\documentstyle[sprocl]{article}

\def\Journal#1#2#3#4{{#1} {\bf #2}, #3 (#4)}
\def\NPB{{\em Nucl. Phys.} B}
\def\PLB{{\em Phys. Lett.}  B}
\def\PRL{\em Phys. Rev. Lett.}
\def\PRD{{\em Phys. Rev.} D}

\def\be{\begin{equation}}
\def\ee{\end{equation}}
\def\bea{\begin{eqnarray}}
\def\eea{\end{eqnarray}}
\begin{document}
\title{PARITY ODD BUBBLES IN HOT QCD}
\author{DMITRI KHARZEEV,}
\address{RIKEN BNL Research Center,\\ Brookhaven National Laboratory,\\
Upton, New York 11973-5000, USA\\}
\author{ROBERT D. PISARSKI,} 
\address{Department of Physics,\\ Brookhaven National Laboratory,\\
Upton, New York 11973-5000, USA\\}
\author{MICHEL H. G. TYTGAT}
\address{Service de Physique Th\'eorique, CP 225,\\ 
Universit\'e Libre de Bruxelles,\\
Blvd. du Triomphe, 1050 Bruxelles, Belgium}
\maketitle\abstracts{We consider the topological
susceptibility for an $SU(N)$ gauge theory in the
limit of a large number of colors, $N \rightarrow \infty$.
At nonzero temperature, the behavior of the topological
susceptibility depends upon the order of the deconfining
phase transition.  The most interesting possibility
is if the deconfining transition, at $T=T_d$, is of second order.
Then we argue that Witten's relation implies that the topological
susceptibility vanishes in a calculable fashion at $T_d$.
As noted by Witten, this implies that for sufficiently light
quark masses, metastable states which act like
regions of nonzero $\theta$ --- parity odd bubbles --- can arise
at temperatures just below $T_d$.  Experimentally, 
parity odd bubbles have dramatic signatures:
the $\eta'$ meson, and especially the $\eta$ meson, 
become light, and are copiously produced.  Further, 
in parity odd bubbles, processes which are normally forbidden, such as 
$\eta \rightarrow \pi^0 \pi^0$, are allowed.
The most direct way to detect parity violation is by 
measuring a parity
odd global asymmetry for charged pions, which we define.}

\section{Introduction}

In this paper we give a pedagogical introduction to recent
work of ours.\cite{pap}  We consider an $SU(N)$ gauge theory
in the limit of a large number of colors, $N\rightarrow \infty$.
This is, of course, a familiar limit.\cite{largeN}  
We use the large $N$ expansion to investigate the behavior
of the theory at nonzero temperature, especially for the
topological susceptibility.
The results depend crucially upon the order of the deconfining
phase transition; if it is first order, nothing very interesting
happens.  If the deconfining transition is of second order, however,
the topological susceptibility vanishes in a calculable fashion.

This implies that metastable states, which act like regions with
nonzero $\theta$, can appear.  Parity is spontaneously broken
in such parity odd bubbles, and produces novel physics.  The 
$\eta$ meson becomes very light,
at most a few hundred $MeV$, and so is easily produced.  As
parity is broken, the $\eta$ can decay into two pions, 
instead of the usual three.  
We also propose a global variable which can be used to measure
an asymmetry in parity.

\section{Large $N$}

Holding the number of fermion flavors 
fixed as $N \rightarrow \infty$, the large $N$ limit is very much a
gluonic limit,
as the $\sim N^2$ gluons totally dominate the $\sim N$ quarks.
This is the basis for most of the conclusions which we can draw
at large $N$: given what happens to the gluons, what happens to
the quarks follows almost immediately.

The standard assumptions at large $N$ is that the physics is
like that for $N=3$: confinement occurs, so at low temperature we
can speak entirely of mesons and glueballs.  Their masses are
assumed to be of order one as $N \rightarrow \infty$; as usually
occurs in any large $N$ limit, interactions between either
mesons and/or glueballs are suppressed by
powers of $1/N$.  

It is also natural to assume that the degeneracy of mesons and
glueballs is of order one.  This is, after all, what we mean
by confinement: all trace of the color indices disappear, leaving
bound states which are characterized only by spin, parity, etc.
Thus in the low temperature phase at a temperature
$T$, since each meson or glueball 
has a free energy which is $\sim N^0 T^4$, the total free energy
is also of order one.  In contrast, in the deconfined
phase at high temperature, the free energy is of order $\sim N^2$.
As pointed out by Thorn,\cite{thorn}
this allows us to use the free energy itself as an order parameter
for the phase transition: the transition, at a temperature
$T_d$, occurs when the term in the free energy $\sim N^2$ turns
on.\cite{rplargeN,otherlargeN}  

Rigorously, the true order 
parameter for the deconfining phase transition 
is associated with the spontaneous breaking of the
global $Z(N)$ symmetry above $T_d$;
this symmetry becomes $O(2)$ as $N\rightarrow \infty$.

We also make a further assumption, namely that {\it any} other
phase transitions occur at the same time as deconfinement, at $T_d$.
We have no proof of this statement, although we suspect that a
proof can be constructed in the limit of large $N$.
Nevertheless, it strains credulity to image that at
the point where this huge increase in the free energy occurs, that
that alone doesn't force any other phase transitions in the theory.

In particular, assume that we couple massless quarks to
the gluons.  At zero temperature, it is known that the quarks' chiral
symmetry must be broken in the familiar pattern, to a diagonal
subgroup of flavor.\cite{coleman}  Then we assume that the chiral
symmetry is restored at a temperature $T_\chi$, with
$T_\chi = T_d$.  

We take the scale of the deconfining transition to be the same
as for the glueball masses; thus $T_d$ is of order one as 
$N\rightarrow \infty$.  This turns out to be a remarkably powerful
assumption.  Consider the large $N$ limit of a theory without
confinement, such as a $N$-component vector with coupling
$g^2$, holding $g^2 N$ fixed as $N\rightarrow \infty$.  We assume
that the masses of the fields are of order one.  Then the only
way for a transition to occur in what is, after all, free field
theory, is to go to temperatures which grow with $N$; a simple
one loop estimate gives $T_\chi \sim 1/\sqrt{g^2} \sim \sqrt{N}$
(this is also the scale of $f_\pi$, which is natural).
What happens in a confining theory is far more dramatic: the
transition occurs at temperatures of order one, not $\sim \sqrt{N}$.
This implies that the hadronic phase is ``cold'' at large $N$:
interactions are small, so that effects from the thermal bath,
such as the loss of manifest Lorentz invariance, can be neglected.

The crucial thing which we do not know about the large $N$ limit
is the order of the deconfining phase transition.  (The effect
of quarks can be neglected, since the gluons dominate the
free energy above $T_d$; this will be elaborated later.)
For this we must look to the lattice, which as always provides
the true intellectual basis for our understanding.

In the early days of Monte Carlo simulations on the lattice,\cite{Nfour}
it was generally agreed that the deconfining phase
transition is of first order when $N=4$.  
It is not clear, however, if these simulations are definitive.
In particular, they were done at $n_t=4$, where
$n_t$ is the number of steps in the imaginary time direction.
For the
standard Wilson action, at this value of $n_t$ there is a bulk
transition close to the finite temperature transition.
The bulk transition can be avoided by going to larger values of
$n_t$, but at the time this was computationally
difficult to do.  Recently,
however, Ohta and Wingate\cite{ohta} have computed for $N=4$ and $n_t=6$;
they find that the strong first order transition at $n_t = 4$
is gone for $n_t=6$.  Of course, to really establish that there
is a true second order phase transition is a difficult matter,
requiring lengthy study.  But these results do suggest that
it may be hasty to conclude from $n_t = 4$
that the deconfining phase transition is of first order.

There are also results on the large $N$ limit of gauge theories
on the lattice.\cite{reduced}   Such reduced models appear to
reliably predict the ratio of the critical temperature
to the square root of the string tension.  They predict a first
order transition, but only under the technical assumption that
the coupling between spacelike plaquettes can be neglected.  It
is not apparent to us how strong this assumption is.

Previously, Pisarski and Tytgat\cite{rdptyt} suggested that
the large $N$ deconfining phase transition is of second order.
Their argument was rudimentary: the easiest way to understand
why the deconfining transition is {\it weakly} first order
for $N=3$ is if the large $N$ expansion is a good approximation,
and if the transition is of second order for $N=\infty$.  Then the
cubic invariant, which drives the transition first order at $N=3$,
is suppressed by $\sim 1/N$.  Of course the first assumption is
rather strong: perhaps the large $N$ expansion is not
a good guide to thermodynamic properties.

In the following 
we assume that the deconfining transition is of second order
for all $N \geq 4$, but comment upon how our results change if
the transition is of first order.

The principal object we are interested in is the topological
suceptibility.\cite{witten1,meggiolaro}  From the topological charge density,
\begin{equation}
Q(x) = (g^2/32\pi^2) tr
(G_{\alpha \beta} \widetilde G^{\alpha \beta}) = \partial_\alpha K^\alpha \; .
\label{eq:ea}
\end{equation}
The current $K^\alpha$ is gauge dependent.  The topological
susceptibility is the two point function of $Q$,
\begin{equation}
\lambda_{YM}(T) \equiv \partial^2 F(\theta, T)/\partial\theta^2
= \int d^4 x \, Q(x) Q(0) \; ;
\label{eq:eb}
\end{equation}
$F(\theta,T)$ is the free energy, and the 
$\theta$ parameter is conjugate to $Q$.  
At zero temperature, the free energy reduces to the energy,
$F(\theta,0)=E(\theta)$.  

Since $Q$ is a total derivative, $\lambda_{YM}(T)$ vanishes
order by order in perturbation theory.  It receives contributions
entirely from nonperturbative effects, such as instantons.
The action of a single instanton with fixed scale size
is $8\pi^2/g^2$.  In the large $N$ limit, 
$g^2 N$ is held fixed as $N\rightarrow \infty$, so the contribution of
an instanton to the topological susceptibility is
$\lambda_{YM}(T) \sim exp(- a N)$, with $a = 8 \pi^2/(g^2 N)$.  
Thus the contribution of instantons vanishes exponentially in
the large $N$ limit.  This naive argument assumes that the integral
over instanton scale size is well behaved.  This is certainly
true in the limit of high temperature; then the theory is weakly
coupled, and instantons are suppressed by the Debye screening
of electric fluctuations.  
This naive picture was verified, at all temperatures,
by Affleck in a soluble asymptotically
free theory, the $CP^N$ model in $1+1$ dimensions\cite{affleck}.  

Thus at large $N$, in the deconfined phase
the topological susceptibility is exponentially
small in $1/N$, and so essentially vanishes.  At zero
temperature, Witten\cite{witten1} suggested that instead
of semiclassical fluctuations, that quantum fluctuations generate
a nonzero value for $\lambda_{YM}(0) \sim N^0$.\cite{witten1}
It is natural to assume that the topological susceptibility
is $\sim N^0$ throughout the deconfined phase, and changes 
to $\sim exp(-a N)$ at the deconfining phase transition.
This was previously argued by Affleck\cite{affleck} and
by Davis and Matheson.\cite{davis}

How it changes depends upon the order of the phase transition.
If the deconfining transition is of first order, then as the
hadronic phase is ``cold'', the most natural possibility
is that the topological susceptibility is essentially constant
in the hadronic phase, and changes discontinuously to zero
at $T_d$. Recent lattice results\cite{DiGiacomo} in full QCD with four 
flavors indicate a sharp drop for the topologocal susceptibility 
across the phase transition, and thus seem to support this conjecture.  

If the deconfining transition is of second order, a more
extended analysis is necessary.\cite{pap}  Generalizing
the results of Witten\cite{witten1}
and Veneziano\cite{veneziano} to nonzero temperature, and
using results on the anomalous couplings of mesons,\cite{anom}
we find that the free energy depends upon $\theta$ as
\begin{equation}
F(\theta,T) \;
\sim_{\!\!\!\!\!\!\!\!\!_{ T\rightarrow T_d^-}} 
(1 + c \, \theta^2) (T_d - T)^{2 - \alpha} \;\; .
\label{eq:ec}
\end{equation}
Here $\alpha$ is the critical exponent for the deconfining phase
transition, $\alpha \approx -.013$.
We then use Witten's\cite{witten1} formula for the $\eta'$ mass
to conclude that the $\eta'$ mass vanishes at $T_d$,
\begin{equation}
m^2_{\eta '}(T) = \frac{4 N_f}{f^2_\pi(T)} \; \lambda_{YM}(T) \;
\sim_{\!\!\!\!\!\!\!\!\!_{ T\rightarrow T_d^-}} 
\;(T_d - T)^{1 - \alpha} \; .
\label{eq:ed}
\end{equation}
Implicitly, we have used the fact that the hadronic phase is cold,
so that zero temperature formulas, such as (\ref{eq:ed}), generalize
trivially.  

\section{Parity odd bubbles}

We now use this result on the $\eta'$ mass to investigate the
nature of the theory in the hadronic phase, just below $T_d$.
At zero temperature, a successful phenomenology of the $\eta'$
was developed with a chiral lagrangian formalism.\cite{witten1,witten2}
For $N_f$ flavors, a $U(N_f)$ matrix $U$ is introduced,
satisfying $U^\dagger U = 1$.  $U$ describes the $N_f^2 - 1$
pions and the $\eta'$.  The effects of the anomaly are
represented solely by a mass term for the $\eta'$,
$(tr \; ln \; U)^2$.

This is stark contrast to how effects of the anomaly due to
instantons are included.  Consider a linear sigma model
with a field $\Phi$.  Then the effects of the anomaly enter exclusively
through a term $\sim det(\Phi)$.  As in the nonlinear sigma
model, when the chiral symmetry is spontaneously broken, this
term generates a mass for the $\eta'$.  However, in the nonlinear
sigma model, at large $N$, 
there is {\it only} a mass term for the $\eta'$; four point
interactions between $\eta'$'s are induced by the anomaly,
but are suppressed by higher powers of $1/N^2$.  As emphasized
by Witten,\cite{witten2} a term $\sim det(\Phi)$ violates this
large $N$ counting.  This is subject to the trivial qualification that
$N_f \geq 4$, so that there are quartic interactions between the
$\eta'$'s.

For $U$ fields which are constant in spacetime, the potential
for $U$ is
\begin{equation}
V(U) = \frac{f_\pi^2}{2}
\left( tr\left( M(U + U^\dagger) \right) 
- a (tr \; ln \; U - \theta)^2 \right) \; ;
\label{eq:ee}
\end{equation}
The pion decay constant $f_\pi = 93\ MeV$, while 
$M$ is the quark mass matrix.
When $M=0$, $m^2_{\eta '} \sim a$, so $a \sim \lambda_{\eta'}^2/N$.  

Taking $M_{i j} = \mu_i^2 \delta^{i j}$,
any vacuum expectation value (v.e.v) of $U$ can be assumed to be diagonal, 
$U_{i j} = e^{i \phi_i} \delta^{i j}$.  The potential reduces to
\begin{equation}
V(\phi_i) = f_\pi^2 \left(
- \sum_{i} \mu_i^2 \; cos(\phi_i)
+ \frac{a}{2} (\sum_i \phi_i - \theta )^2 \right) \; .
\label{eq:ef}
\end{equation}
This is minimized for
\begin{equation}
\mu_i^2 \; sin(\phi_i) + a (\sum \phi_i - \theta) = 0 \; .
\label{eq:eg}
\end{equation}
Note that as  $\sum \phi_i$ arises from
$tr \, ln \, U$, it is defined modulo $2 \pi$.

Previously, several authors studied how the v.e.v.'s of the
$\phi$'s change as a function of $\theta$.\cite{largeNphen1,largeNphen2}
In the present work, we consider $\theta=0$, but consider how
the $\mu_i$ and $a$ change with temperature.
Witten\cite{witten2} pointed out that when the anomaly term $a$ becomes
small, metastable states in the $\phi$'s can arise.  
From our arguments in the previous section, this happens naturally
if the phase transition at large $N$ is of second order.

The presence of these metastable states can be easily understood
for a single flavor, as discussed by Witten\cite{witten2} (for a 
recent discussion see\cite{shifman,hz}). 
From (\ref{eq:eg}), the v.e.v. arises from a balance between
a term $\sim sin(\phi)$ and a term $\sim \phi$.  For large
$a$, the term linear in $\phi$ wins, and there is no 
possibility for a metastable point.  Now consider the
opposite limit, of vanishing $a$: 
then there automatically other solutions besides $\phi=0$,
$\phi = 2 \pi, 4 \pi$, etc.  These solutions are equivalent
to the trivial vacuum, and so there is nothing new.  But
for small values of $a$, the term linear in $a$ will only
move the stationary point a little bit from $2 \pi$, $4\pi$, etc.
Because $a$ is nonzero, they will become metastable, distinct
from the usual vacuum.  

From (\ref{eq:eg}), these states will act like regions of nonzero
$\theta$.  Parity and CP are both violated
spontaneously in such a region.  

The condition for metastable states to arise with more than
one flavor is not apparent, and a new result of our analysis.\cite{pap}
It is easiest understood by analogy.  At zero temperature, 
and nonzero $\theta$, if any quark mass vanishes, the $\theta$
parameter can be eliminated by a chiral rotation through that
quark flavor.  Thus it is the lightest quark mass which controls
$\theta$ dependence.  We found a similar phenomenon for metastable
states: they only occur when the anomaly term is small relative
to the lightest quark masses.  

This means that metastable states only arise when the anomaly
term becomes very small.  At zero temperature, the anomaly term
is on the order of the strange quark mass.  The previous argument
indicates that it must become on the order of the up and down
quark masses.  Putting in the numbers, we find that metastable
states only arise when the anomaly term becomes on the order of
$1\%$ of its value at zero temperature.  Clearly this is a strong
variation of the topological susceptibility with temperature;
nevertheless, it is interesting to investigate the possible
implications for phenomenology.

Most notably,
when the anomaly term $a$ becomes small, there is maximal violation
of isospin.\cite{largeNphen1,ren1,small1,small2}  At zero temperature, the 
nonet of pseudo-Goldstone bosons --- the $\pi$'s, $K$'s, $\eta$,
and $\eta'$, are, to a good approximation, eigenstates of $SU(3)$
flavor.  It is not often appreciated, but this is really due to
the fact that the anomaly term is large, splitting off the $\eta'$
to be entirely an $SU(3)$ singlet.  When the anomaly term becomes
small, however, while the charged pseudo-Goldstone bosons remain
approximate eigenstates of flavor, the neutral ones do not.
Without the anomaly, the $\pi^0$ becomes pure $\overline{u} u$,
the $\eta$ pure $\overline{d} d$, and the $\eta'$ pure
$\overline{s} s$.  Consequently, these three mesons become light.
This is especially pronounced for the $\eta$, as it sheds all
of its strangeness, to become purely $\overline{d} d$.
Thus the $\eta$ and $\eta'$ would be produced copiously,
and would manifest itself in at least two ways.  First,
light $\eta$'s and $\eta'$'s decay into two photons, and
so produce an excess at low momentum.  Secondly, these
mesons decay into pions, which would be seen in Bose-Einstein
correlations\cite{bose}. Further, through Dalitz decays, 
the enhanced production 
of  $\eta$'s and $\eta'$'s will enhance the yield of low mass dileptons
\cite{small2}. 

This maximal violation of isospin is true whenever the anomaly
term becomes small.  There are other signals which only
appear when parity odd bubbles are produced.  Since parity
is spontaneously violated in such a bubble, various decays,
not allowed in the parity symmetric vacuum, are possible.
Most notably, the $\eta$ can decay not just to three pions,
as at zero temperature, but to two pions.  Because of the kinematics,
in a parity odd bubble, $\eta \rightarrow \pi^0 \pi^0$ is allowed,
but $\eta \rightarrow \pi^+ \pi^-$ is not.

There is another measure of how parity may be violated.  We first
argue by analogy.  Consider propagation in a background magnetic
field.  As charged particles propagate in the magnetic field,
those with positive charge are bent one way, and those with
negative charge, the other.  This could be observed by measuring
the following variable globally, on an event--by--event basis:
\begin{equation}
{\cal P} = \sum_{\pi^+\pi^-}
\frac{(\vec{p}_{\pi^+} \times \vec{p}_{\pi^-})\cdot \vec{z}}
{ |\vec{p}_{\pi^+}| |\vec{p}_{\pi^-}|} \; ;
\label{eq:eh}
\end{equation}
here $\vec{z}$ is the beam axis, and $\vec{p}$ are the three momenta
of the pions.

If the quarks were propagating through a background
chromo-magnetic field, then ${\cal P}$, which is like handedness
in jet physics,\cite{hand} is precisely the right quantity.
However, a parity odd bubble is
not directly analogous to a background chromo--magnetic field: $\pi^+$'s
and $\pi^-$'s propagate in a region with constant but
nonzero $\phi$ in the same fashion.  Consider, however,
the edge of the parity odd bubble: in such a region, 
$U^\dagger \partial_\mu U$ is nonzero, and does rotate $\pi^+$
and $\pi^-$ in opposite directions.  Thus it is the edges
of parity odd bubbles which contribute to the parity
odd asymmetry of (\ref{eq:eh}).  Purely on geometric grounds,
this suggests that a reasonable
estimate for the maximal value of ${\cal P}$ is on the order of
a few percent.

We conclude by noting that what appears to be a rather
technical subject --- the $\theta$ dependence of the free energy ---
is related to interesting and novel experimental signatures
in heavy ion collisions.  Within our assumptions, we
find that parity odd bubbles only arise very near the point of
the phase transition.  This is very much tied to the fact that
we limit ourselves to an analysis at large $N$.  For finite $N$,
it is a long standing question of how to reconcile the known limit
at large $N$ with periodicity in $\theta$, with period $2 \pi$.
A probable solution involves ``glued'' potentials,
which are a sum of cosines; see\cite{shifman,hz}
and recently.\cite{theta}  The precise
form of the potential at finite $N$ could dramatically alter
our results, and, as discussed by Halperin and Zhitnitsky,\cite{theta} 
make the emergence of parity odd bubbles far more likely.


\section*{References}
\begin{thebibliography}{99}
\bibitem{pap}D. Kharzeev, R. D. Pisarski, and M. H. G. Tytgat,
\Journal{\PRL}{81}{512}{1998}.
\bibitem{largeN}
G. 't Hooft, \Journal{\NPB}{72}{461}{1974};
E. Witten, \Journal{\NPB}{160}{57}{1979};
A. V. Manohar, {\em hep-ph/9802419};
M. Teper, \Journal{\PLB}{397}{223}{1997}; {\em hep-lat/9804008}.
\bibitem{thorn}
C. B. Thorn, \Journal{\PLB}{99}{458}{1981}.
\bibitem{rplargeN}
R. D. Pisarski, \Journal{\PRD}{29}{1222}{1984}.
\bibitem{otherlargeN}
J. J. Atick and E. Witten, \Journal{\NPB}{310}{291}{1988};
J. Polchinski, \Journal{\PRL}{68}{1267}{1992};
T. H. Hansson and I. Zahed, \Journal{\PLB}{309}{385}{1993}.
\bibitem{coleman}S. Coleman and E. Witten,
\Journal{\PRL}{45}{100}{1980}.
\bibitem{Nfour}
G. G. Batrouni and B. Svetitsky, 
\Journal{\PRL}{52}{2205}{1984};
A. Gocksch and M. Okawa, 
\Journal{\PRL}{52}{1751}{1984};
F. Green, \Journal{\PRD}{29}{2986}{1984};
J. F. Wheater and M. Gross, 
\Journal{\PLB}{144}{409}{1984};\Journal{\NPB}{240}{253}{1984}.
\bibitem{reduced}
M. Billo, M. Caselle, A. D'Adda, and S. Panzeri,
{\em Int. J. Mod. Phys. A} {\bf 12}, 1783 (1997).
\bibitem{ohta}S. Ohta and M. Wingate, {\em hep-lat/9808022}.
\bibitem{rdptyt}R. D. Pisarski and M. H. G. Tytgat, {\em hep-ph/9702340}.
\bibitem{witten1}
E. Witten, {\em Nucl. Phys. B} {\bf 156}, 269 (1979). 
\bibitem{veneziano}
G. Veneziano, {\em Nucl. Phys. B} {\bf 159}, 213 (1980).
\bibitem{largeNphen1}
P. Di Vecchia and G. Veneziano, {\em Nucl. Phys. B} {\bf 171}, 253 (1980).
\bibitem{largeNphen2}
P. Di Vecchia, {\em Phys. Lett. B} {\bf 85}, 357 (1979);
P. Di Vecchia, F. Nicodemi, R. Pettorino, and G. Veneziano,
{\em Nucl. Phys. B} {\bf 181}, 318 (1981);
P. Nath and R. Arnowitt, {\em ibid.} {\bf 209}, 234, 251 (1982);
C. Rosenzweig, J. Schechter, and C. G. Trahern,
{\em Phys. Rev. D} {\bf 21}, 3388 (1980).
\bibitem{witten2}
E. Witten, {\em Annals Phys.} {\bf 128}, 363 (1980).
\bibitem{shifman}
N. Evans, S. D. H. Hsu, and M. Schwetz,
{\em Nucl. Phys. B } {\bf 484}, 124 (1997); 
{\em ibid.}, {\bf 494}, 200 (1997);
M. Shifman, {\em Prog. Part. Nucl. Phys.} {\bf 39}, 1 (1997);
I. I. Kogan, A. Kovner, and M. Shifman,
{\em Phys. Rev. D} {\bf 57}, 5195 (1998).
\bibitem{hz}
I. Halperin and A. Zhitnitsky, {\em hep-ph/9707286; hep-ph/9803301}.
\bibitem{affleck}I. Affleck, 
\Journal{\NPB}{162}{461}{1980}; 
{\em ibid.}, {\bf 171}, 420 (1980).
\bibitem{davis}
A. C. Davis and A. M. Matheson, {\em Phys. Lett. B} {\bf 179}, 135 (1986);
{\em Nucl. Phys. B} {\bf 258}, 373 (1985).
\bibitem{DiGiacomo}
P. de Forcrand, M. G. Perez, J. E. Hetrick, and I.-O. Stamatescu,
{\em hep-lat/9802017};
B. Alles, M. D'Elia, A. Di Giacomo, 
and P. W. Stephenson, {\em hep-lat/9808004}.
\bibitem{meggiolaro}
E. Meggiolaro, {\em Zeit. fur Phys.} {\bf 62}, 679, 669 (1994);
{\em ibid.} {\bf 64}, 323 (1994); {\em hep-th/9802114}.
\bibitem{anom}
R.D.~Pisarski, 
in {\em From thermal field theory to neural networks: a day to
remember Tanguy Altherr}, 
edited by P. Aurenche, P. Sorba, and G. Veneziano.
(World Scientific Publishing, Singapore, 1996);
{\em Phys. Rev. Lett.} {\bf 76}, 3084 (1996);
R. Baier, M. Dirks, and O. Kober, {\em Phys. Rev. D} {\bf 54}, 2222 (1996);
R. D. Pisarski, T. L. Trueman, and M. H. G. Tytgat,
{\em ibid.}, {\bf 56}, 7077 (1997).
\bibitem{ren1}
R. D. Pisarski and F. Wilczek, 
{\em Phys. Rev. D} {\bf 29} (1984) 338.
\bibitem{small1}
Z. Huang, {\em Phys. Rev. D} {\bf 49}, 16 (1994).
\bibitem{small2}
J. Kapusta, D. Kharzeev and L. McLerran, 
{\em Phys. Rev. D} {\bf 53}, 5028 (1996); 
Z. Huang and X.-N. Wang, {\em ibid.,} {\bf 53}, 5034 (1996).
\bibitem{bose}
S. E. Vance, T. Cs\"{o}rg\H{o} and D. Kharzeev, {\em nucl-th/9802074; 
Phys. Rev. Lett.}, to appear.
\bibitem{hand}
O. Nachtmann, {\em Nucl. Phys. B} {\bf 127}, 314 (1977); 
A.V. Efremov, {\em Sov. J. Nucl. Phys.} {\bf 28}, 83 (1978);
A. Efremov and D. Kharzeev, {\em Phys. Lett. B} {\bf 366}, 311 (1996).  
\bibitem{theta}
A. V. Smilga, {\em hep-ph/9805214}; E. Witten, {\em hep-th/9807109};
I. Halperin and A. Zhitnitsky, {\em hep-ph/9807335}.
\end{thebibliography}
\end{document}